\documentclass[final,5p,times,twocolumn, 12pt]{elsarticle}
\usepackage{array}
\usepackage{amssymb}
\usepackage{amsmath}
\usepackage{csquotes}
\usepackage[breaklinks=true]{hyperref}

\usepackage{subcaption}
\newcommand{\labelphantom}[1]{%
{\phantomsubcaption%
\label{#1}}%
}%
\makeatletter
\renewcommand\p@subfigure{\thefigure-}
\makeatother

\journal{Fusion Engineering and Design}

\begin{document}

\begin{frontmatter}



\title{Electromagnetic System Conceptual Design for a Negative Triangularity Tokamak} 



\author[Columbia]{S. Guizzo}
\author[NextStep]{M.A. Drabinskiy}
\author[Columbia]{C. Hansen}

\author[NextStep]{A.G. Kachkin}
\author[NextStep]{E.N. Khairutdinov}
\author[Columbia]{A. O. Nelson}

\author[NextStep]{M.R. Nurgaliev}

\author[Columbia]{M. Pharr}

\author[NextStep]{G.F. Subbotin}

\author[Columbia]{C. Paz-Soldan}

\affiliation[Columbia]{organization={Columbia University},
  city={New York City},
  state = {New York},
  postcode={10027},
  country={USA}}

\affiliation[NextStep]{organization={Next Step Fusion},
  city={Luxembourg}}

%% Abstract
\begin{abstract}
Negative triangularity (NT) tokamak configurations have several key benefits including sufficient core confinement, improved power handling, and reduced edge pressure gradients that allow for edge-localized mode (ELM) free operation. We present the design of a compact NT device for testing sophisticated simulation and control software, with the aim of demonstrating NT controllability and informing power plant operation. The TokaMaker code is used to develop the basic electromagnetic system of the $R_0$ = 1 m, $a$ = 0.27 m, $B_t$ = 3 T, $I_p$ = 0.75 MA tokamak. The proposed design utilizes eight poloidal field coils with maximum currents of 1 MA to achieve a wide range of plasma geometries with  $-0.7 < \delta < -0.3$ and $1.5 < \kappa < 1.9$. Scenarios with strong negative triangularity and high elongation are particularly susceptible to vertical instability, necessitating the inclusion of high-field side and/or low-field side passive stabilizing plates which together reduce vertical instability growth rates by $\approx$75\%.  Upper limits for the forces on poloidal and toroidal field coils are predicted and mechanical loads on passive structures during current quench events are assessed.  The 3 T on-axis toroidal field is achieved with 16 demountable copper toroidal field coils, allowing for easy maintenance of the vacuum vessel and poloidal field coils. This pre-conceptual design study demonstrates that the key capabilities required of a dedicated NT tokamak experiment can be realized with existing copper magnet technologies. 
\end{abstract}



%% Keywords

\end{frontmatter}



%%%%%%%%%%%%%%% INTRODUCTION %%%%%%%%%%%%%%
%%%%%%%%%%%%%%%%%%%%%%%%%%%%%%%%%%%%%%%%%%%

\section{Introduction}
\label{sec:intro}
Negative triangularity (NT) tokamak scenarios are a topic of recent interest in the fusion community due to several intrinsic benefits over traditional positive triangularity H-mode operation. Simulation and experiment have demonstrated the impressive performance of the NT core scenario, attributed to a reduction in core turbulent transport that enhances energy confinement \cite{paz-soldan_simultaneous_2024,giannatale_system_2024, merlo_effect_2023}. The NT cross-section also places strike-points on the low-field side of the device at a larger major radius, resulting in a larger wetted area for heat deposition \cite{Medvedev2015}. Furthermore, plasmas with NT shaping reach high core pressures while maintaining a reduced edge pressure gradient, preventing the occurence of destructive edge-localized modes (ELMs) \cite{nelson_characterization_2024, nelson_robust_2023, nelson_prospects_2022,saarelma_ballooning_2021}. As a result, negative triangularity is currently being considered as a candidate tokamak concept for a fusion pilot plant \cite{Medvedev2015,collaboration_manta_2024, wilson_characterizing_2025}. 

Many existing and planned tokamak experiments are challenged by NT operation due to geometric constraints. Owing to their flexibility, the DIII-D \cite{thome_overview_2024, paz-soldan_simultaneous_2024, marinoni_diverted_2021}, TCV\cite{balestri_experiments_2024, pochelon_energy_1999, Camenen2007}, and to a more modest degree ASDEX-Upgrade \cite{aucone_experiments_2024, merle_pedestal_2017} tokamaks have successfully explored the NT operational space and developed the physics basis necessary to promote interest in the regime. Recently, NT has been achieved for the first time at low aspect ratio in the MAST-U tokamak \cite{nelson_mastu_nt}. Such experiments have motivated the design and construction of further tokamaks capable of testing the NT physics basis. This includes the SMall Aspect Ratio Tokamak (SMART), a spherical tokamak targeting a wide range of plasma geometries with $-0.5 < \delta < 0.5$, currently in development at the University of Seville \cite{doyle_2021, mancini_mechanical_2021, SEGADOFERNANDEZ2023113832}.

%in addition to physics basis, need to develop key technologies required to run an fpp
%need less user input
%disruptions are worse
%private interest in this problem
%need dedicated facilities to test

These user-facilities are well-suited for scientific exploration and conduct studies in a diverse range of physics and fusion-relevant topics. However, it is equally important to develop key technologies and software necessary for the realization of fusion reactors. Pilot plants and subsequent devices will need to operate with less user input than current experimental tokamaks. Furthermore, off-normal events such as unmitigated disruptions could be very costly in larger scale, high-field devices \cite{LEHNEN201539, maris_impact}. 
 Automated control of plasma instabilities and real-time prediction and response is a topic of recent interest, particularly in the private sector, and experimental tokamaks are required to test and validate such systems.

%These user-facilities are well-suited for scientific exploration and conduct studies in a diverse range of physics and fusion-relevant topics.  However, as private interest in fusion grows \cite{fusion_company}, there is increasing demand for test devices dedicated to the piloting of commercial fusion technologies and software.  Fusion pilot plants and subsequent devices will need to operate with less user input than current experimental tokamaks .  Furthermore, off-normal events such as unmitigated disruptions could be very costly in larger scale, high-field devices \cite{LEHNEN201539, maris_impact}. Automated control of plasma instabilities and real-time prediction and response is a topic  which require dedicated experiments to test and validate 

%Next Step Fusion is a technology company developing machine learning algorithms to aid in fusion plasma simulation, control, and behavior prediction, which are topics of significant relevance in the commercialization of fusion energy.

%Fusion pilot plants and subsequent devices will need to operate with less user input than current experimental tokamaks .  Furthermore, off-normal events such as unmitigated disruptions could be very costly in larger scale, high-field devices \cite{LEHNEN201539, maris_impact}.  Automated control of plasma instabilities and real-time prediction and response is a topic well-suited for the application of advanced machine learning algorithms \cite{vega_disruption_2022, degrave_magnetic_2022}.

\begin{figure}
    \centering
    \includegraphics[width=1\linewidth]{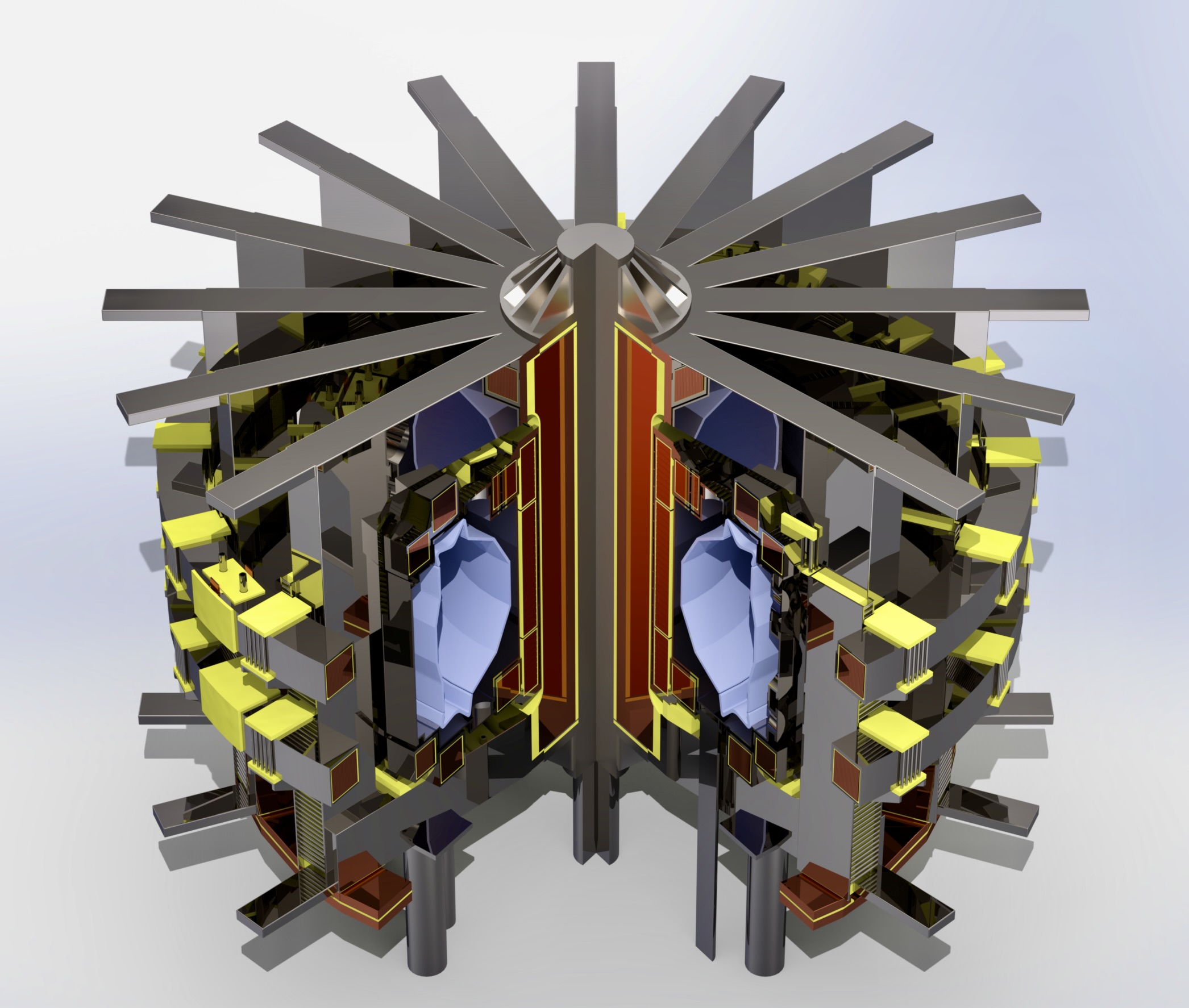} 
    \caption{CAD visualization of the negative triangularity tokamak (NTT).} 
    \label{fig:NTT_3D}%
\end{figure}

In this work, we present the design of a compact experimental tokamak to pilot simulation and control software. The device is targeting NT operation due to its promise as a regime of interest for fusion pilot plants. Furthermore, NT configurations are prone to vertical instability, providing the ideal test-bed for control algorithms \cite{song_impact_2021, guizzo_assessment_2024}. The Negative Triangularity Tokamak (NTT) depicted in figure~\ref{fig:NTT_3D} has major radius $R_0$ = 1 m, on-axis toroidal magnetic field $B_t$ = 3 T, and plasma current near $I_p$ = 750 kA, all achieved with existing copper magnet technology. The device parameters are summarized in table~\ref{table:machine_params}.  The conventional aspect ratio and high magnetic field of the device are chosen to provide the best extrapolation to fusion pilot plant scenarios. The conceptual design and analysis of the NTT was conducted using the TokaMaker Grad-Shafranov solver in the OpenFUSIONToolkit (OFT) \cite{hansen_2023}.

\newcolumntype{M}[1]{>{\centering\arraybackslash}m{#1}}

\begin{center}
\begin{table}
\centering
\begin{tabular}{ | M{4.0cm} | M{1.5cm} | M{1.5cm}|} 
\hline
Parameter & Symbol &  Value \\
\hline
Major radius & $R_0$ & 1 m \\
\hline
Minor radius & $a$ & 0.27 m \\
\hline
Plasma current & $I_\mathrm{p}$ & 750 kA \\
\hline
Toroidal magnetic field & $B_\mathrm{t}$ & 3 T \\
\hline
Pulse length & $\tau_\mathrm{pulse}$ & 10 s\\
\hline
\end{tabular}
\caption{Target parameters for the NT tokamak design.}
\label{table:machine_params}
\end{table}
\end{center}

\vspace{-43 pt}

In section~\ref{sec:pf_coils}, we describe the design of the poloidal field (PF) coil set and resulting domain of accessible plasma geometries.  We also discuss the inclusion of divertor coils to improve power handling prospects via strike point sweeping. In section~\ref{sec:vert_stability}, we employ linear vertical stability calculations to inform operational limits and motivate the inclusion of passive stabilizing plates optimized for vertical stability. We calculate the maximum predicted forces on the electromagnets and the expected forces on passive conducting structures during a current quench to assess the resilience of the design to mechanical stresses in section~\ref{sec:force}. An overview of the preliminary design of the NTT magnetic system is provided in section~\ref{sec:mag_syst}. Finally, in section~\ref{sec:perf_analysis}, we analyze a Plasma OPerational CONtour (POPCON) to estimate overall device performance \cite{houlberg_contour_1982}.

%\clearpage
%\newpage

%%%%%%%%%%%%%%%%%%%%%%%%%%%%%%%%%%%%%%%
%%%%%%%%%%%%% SECTION ONE %%%%%%%%%%%%%
%%%%%%%%%%%%%%%%%%%%%%%%%%%%%%%%%%%%%%%
\section{Optimization of poloidal field coil set}
\label{sec:pfs}
% goal: few PF coils, lots of flexibility in shape
% TF constraint
%final positions in table, along with max current across the delta/kappa scan
%fig 3, shows complete delta scan, different coils stressed for different configurations, without exceeding limits

\label{sec:pf_coils}
\begin{figure}
    \centering
    \includegraphics[width=1\linewidth]{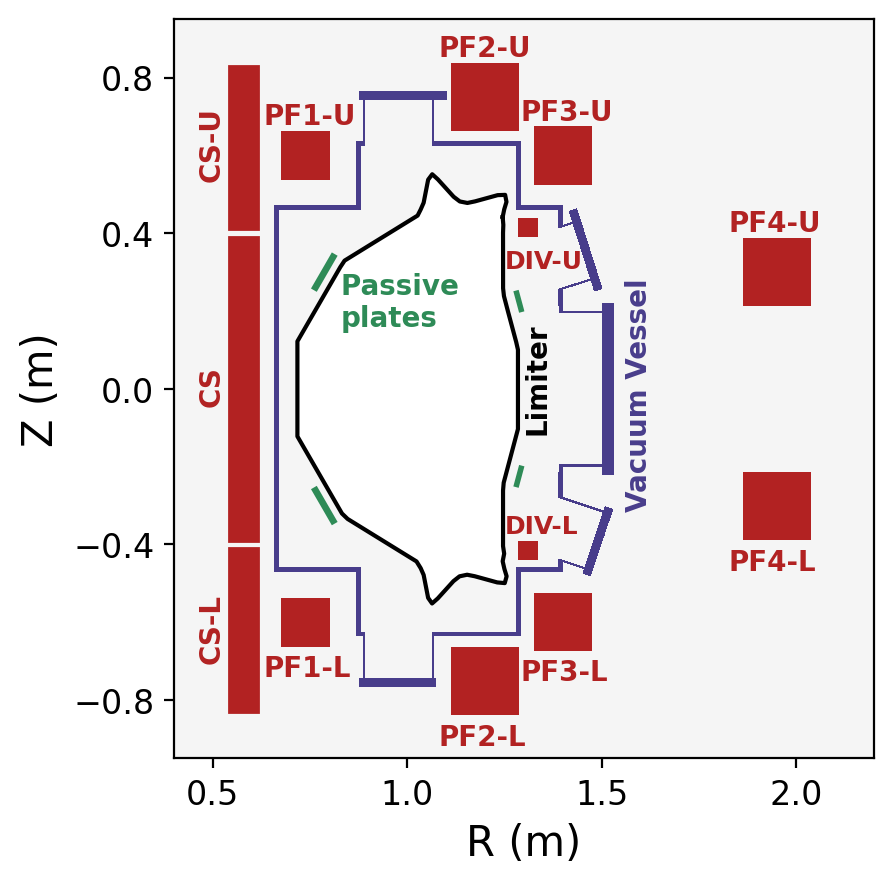} 
    \caption{NT tokamak cross-section with key components labeled.} 
    \label{fig:cross_section}%
\end{figure}

The Negative Triangularity Tokamak is designed to have few poloidal field coils while maintaining a broad range of accessible plasma geometries.  The final configuration includes four sets of up-down symmetric poloidal field coils, depicted in figure~\ref{fig:cross_section}. This coilset can produce plasmas with all combinations of triangularities and elongations in the domain $-0.7 < \delta < -0.3$ and $1.5 < \kappa < 1.9$.  For the remainder of the text, the baseline scenario refers to the plasma geometry with $\delta = -0.5$ and $\kappa = 1.7$, unless otherwise stated. 

%The vacuum vessel of NTT is not the focus of this study; therefore, a vacuum vessel with a rectangular design is considered for reference purposes.

%\begin{comment}
\begin{figure*}[h!]
    \centering
    \labelphantom{fig:delta_scan-a}
    \labelphantom{fig:delta_scan-b}
    \labelphantom{fig:delta_scan-c}
    \includegraphics[width=1\linewidth]{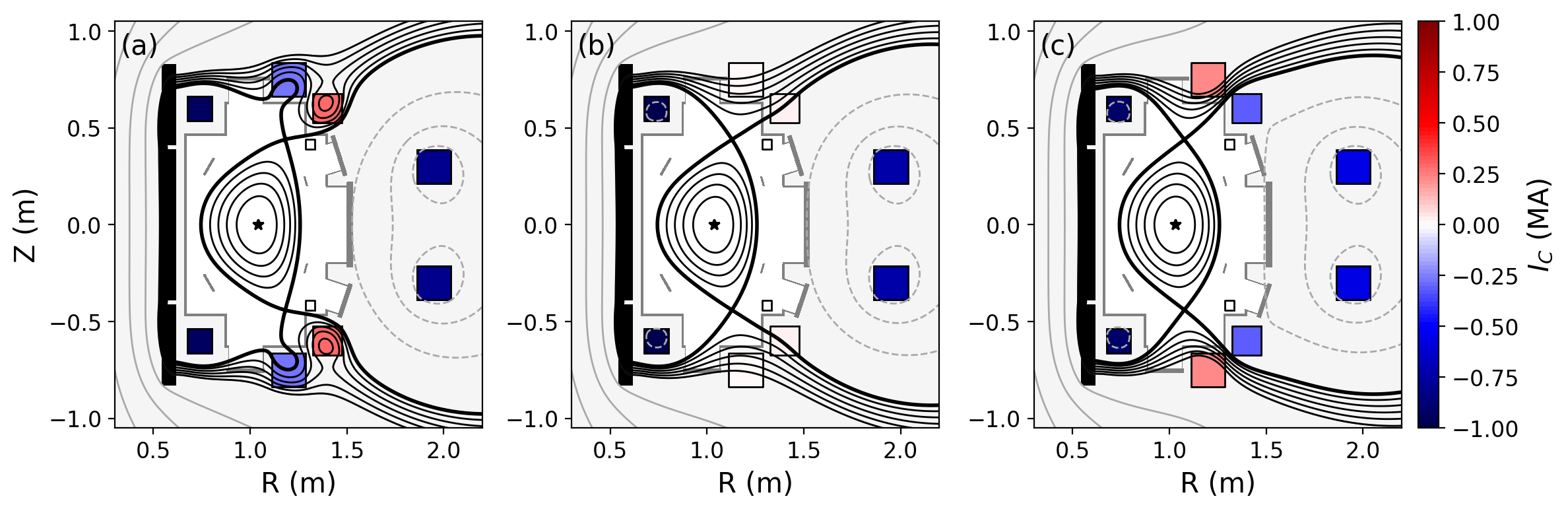} 
    \caption{Baseline scenario with (a) $\delta$ = -0.7, (b) $\delta$ = -0.5, and (c) $\delta$ = -0.3 and corresponding poloidal field coil currents.} 
    \label{fig:delta_scan}%
\end{figure*}
%\end{comment}

The locations and dimensions of the PF coils, as well as the maximum current in each coil across the accessible geometry range for $I_p = 0.75$ MA, are listed in table~\ref{table:pf_currents}. The NTT implements a hybrid topology for the poloidal field coils. PF1, PF2, and PF3 are located inside the toroidal field coils, close to the vacuum vessel, while PF4 is placed outside of the toroidal field coils at $R$ = 1.95 m, necessitating the use of jointed toroidal field coils.  The poloidal field coils are sized according to current density limits as well as device geometry constraints. For example, even though the maximum current in PF1 is high across the shape scan, the dimensions of the coil are limited due to proximity to the central solenoid.

Figure~\ref{fig:delta_scan} shows a triangularity scan at a fixed elongation of $\kappa = 1.7$, with the corresponding poloidal field coil currents.  The current in each segment of the central solenoid is fixed at -3 MA, although self-consistent pulse simulations are in progress to determine full current trajectories. As demonstrated in the figure, PF2 and PF3 share the responsibility of pulling the X-points across the triangularity scan. 
\vspace{-30 pt}

\begin{center}
\begin{table}
\centering
\begin{tabular}{ | M{0.9cm} | M{0.95cm}| M{0.95cm} |M{0.95cm} |M{0.95cm} | M{1.5cm}|} 
\hline
Coil & R(m) & Z(m) & dR (m) & dZ (m)& Maximum Current (MA) \\
 \hline
 PF1  & 0.738 & 0.6 &0.125 & 0.125 &0.98 \\
 \hline
 PF2  & 1.2& 0.75 &0.175 & 0.175 & 0.81\\
 \hline
 PF3 & 1.4 & 0.6 & 0.15 & 0.15 & 0.87 \\
 \hline
PF4 & 1.95 & 0.3 & 0.175 & 0.175 & 0.88 \\
 \hline
\end{tabular}
\caption{Maximum currents in poloidal field coils across a geometry scan of $-0.7 \leq \delta \leq -0.3$ and $1.5 \leq \kappa \leq 1.9$ for $I_{\mathrm{p}}$ = 0.75 MA.}
\label{table:pf_currents}
\end{table}
\end{center}

%\clearpage
%\newpage

\begin{figure*}
    \labelphantom{fig:div_coil-a}
   \labelphantom{fig:div_coil-b}
   \labelphantom{fig:div_coil-c}
   \labelphantom{fig:div_coil-d}
    \centering
    \includegraphics[width=1\linewidth]{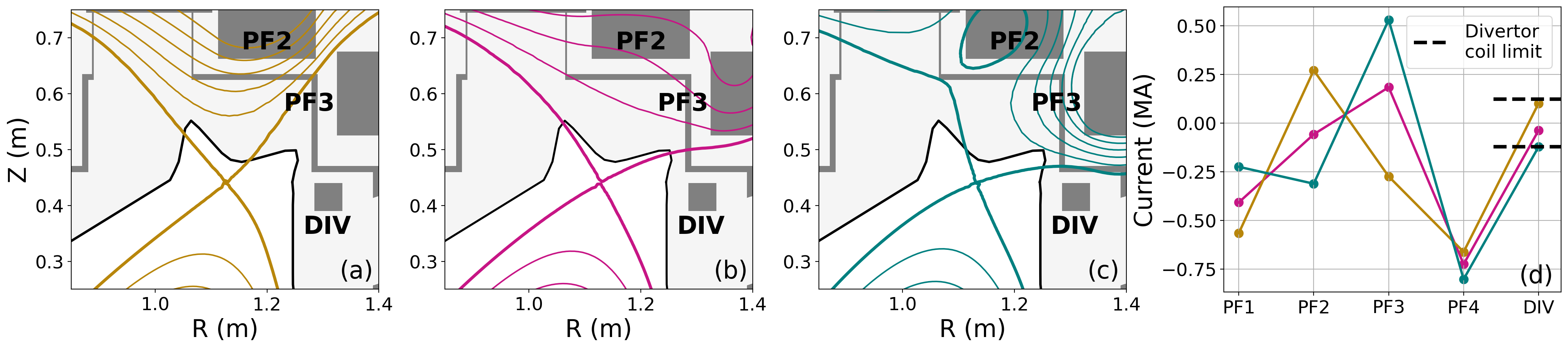} 
    \caption{(a) Left-most, (b) central, and (c) right-most strike point locations with (d) corresponding poloidal field coil and divertor coil currents.} 
    \label{fig:div_coil}%
\end{figure*}

\subsection{Equilibrium Power Handling}
%OUTLINE
%NT has better power handling inherently, but we want to have option for strike point sweeping
% Tested out divertor coils, set same density limit as other PFs
%For baseline, can achieve considerable strike point range while remaining below coil current limits
%inclusion of a HFS divertor coil does not dramatically improve range, nor does it significantly decrease coil current requirements -> not included in final design
%Computed strike line angles for maximum range

One considerable benefit of negative triangularity operation is inherently improved power-handling. Negative triangularity shaping prevents the occurrence of edge-localized modes that cause periodic high heat fluxes on divertor targets during conventional positive triangularity H-mode operation \cite{nelson_characterization_2024, nelson_robust_2023, nelson_prospects_2022,saarelma_ballooning_2021}.  Furthermore, negative triangularity geometries place the strike points, and therefore divertor targets on the outboard side of the device, at a larger major radius, resulting in a larger surface area for heat deposition \cite{Medvedev2015}.  Together, these effects significantly improve power handling prospects in a specialized negative triangularity device, without the need for advanced divertor solutions \cite{advanced_divertor, kuang_arc}.

However, a set of up-down symmetric divertor coils is included in the design of the Negative Triangularity Tokamak to provide the option for strike-point sweeping, which prevents the concentration of heat fluxes on specific portions of the divertor \cite{sparc_div}. The divertor coils are placed on the low-field side of the device, and have current density limits equivalent to the other poloidal field coils. Within these bounds, the divertor coils can precisely control the location of the strike lines, and access a wide range of potential strike point positions.  Figure~\ref{fig:div_coil} shows the maximum strike point range accessible for the baseline configuration with $\delta = -0.5$ and $\kappa = 1.7$, and the corresponding poloidal field and divertor coil currents.  A second set of divertor coils on the high-field side only marginally improves the accessible strike point range, does not reduce any coil current requirements across the sweep, and would interfere with the vertical ports of the vacuum vessel.  As a result, only one pair of divertor coils is included in the final design.

As an additional part of this study, the fully 3D field line incidence angles are calculated for the strike point sweep depicted in figure~\ref{fig:div_coil}.  The incidence angle for the inner strike line remains below $2.5^{\circ}$ throughout the scan, where the angle is measured relative the toroidal direction. However, the outer strike line reaches an incidence angle of $4.0^{\circ}$ at the extremum depicted in figure~\ref{fig:div_coil-c}, potentially increasing the incident heat flux on the divertor substantially. The outer strike line angle is decreased considerably when triangularity is reduced below $\delta = -0.5$.

%\begin{figure}
 %   \centering
 %   \includegraphics[width=0.9\linewidth]{strike_angle.png} 
 %   \caption{(a) Inner and (b) outer strike line angles for strike point sweep depicted in figure~\ref{fig:div_coil}. \textcolor{red}{[OPTIONAL]}} 
 %   \label{fig:strike_angle}%
%\end{figure}

%%%%%%%%%%%%%%%%%%%%%%%%%%%%%%%%%%%%%%%%%
%%%%%%%%%%%%% SECTION THREE %%%%%%%%%%%%%
%%%%%%%%%%%%%%%%%%%%%%%%%%%%%%%%%%%%%%%%%
\section{Vertical stability}
\label{sec:vert_stability}

%OUTLINE
%(1) NT can be more vertically unstable
%(2) with true wall, many configurations ideal wall unstable
%(3) with conformal wall, growth rates are very high, beyond what we expect to be controllable
%(4) passive plates useful for NT, both inboard and outboard side feasible
%(5) Scan indicated most helpful locations for passive plates, give plate specs
%(6) gamma * tau decreases considerably with plates, even for extreme configs
%(7) With true vv + plates, most geometries are in a reasonable gamma*tau space

Negative triangularity shaping can increase drive for the $n=0$ vertical instability \cite{song_impact_2021, Nelson2023, guizzo_assessment_2024}.  In an experiment specifically targeting negative triangularity operation, it is important to assess the ability of conducting structures to provide passive vertical stabilization.  The TokaMaker code \cite{hansen_2023} can compute the growth rate ($\gamma$) of the vertically unstable eigenmode for a given equilibrium as well as the diffusion time for eddy currents in the conducting structures ($\tau_\mathrm{W}$). The resulting feeedback capability parameter $\gamma\tau_{\mathrm{W}}$ can be used to quantify whether a given scenario can be controlled by an active feedback system \cite{freidberg_tokamak_2015}.   

The stainless steel vacuum vessel of the Negative Triangularity Tokamak, as depicted in figure~\ref{fig:cross_section}, has a wall diffusion time of $\tau_\mathrm{W} \approx 5.8$ ms.  The vacuum vessel is not fully conformal to the plasma and features substantial plasma-wall gaps in certain regions.  While this enables a wide range of accessible plasma geometries, the vacuum vessel serves as a poor stabilizer for the $n=0$ vertical mode.  For most scenarios with $-0.7 < \delta < -0.3$ and $1.5 < \kappa < 1.9$, the vertical instability growth rate approaches infinity, indicating that a vertical displacement event would occur on Alfv\'en timescales in experiment.

Due to the infinite vertical instability growth rates for some plasma geometries, further vertical stability studies are performed assuming a 1 cm thick stainless steel vacuum vessel conformal to the plasma boundary at a distance of 10 cm, with $\tau_\mathrm{W} \approx 17.0$ ms.  With this vacuum vessel, all the vertical instability growth rates are large, but finite.  Figure~\ref{fig:vs-a} shows the feedback capability parameter over a scan of accessible elongations and triangularities in the conformal vacuum vessel. Geometries with strong negative triangularity and high elongation have $\gamma \tau_{\mathrm{W}} > 8$, expected to be beyond the capacity of most tokamak vertical control systems \cite{lee_tokamak_2015}. 

\begin{figure}
    \labelphantom{fig:vs-a}
   \labelphantom{fig:vs-b}
   \labelphantom{fig:vs-c}
    \centering
    \includegraphics[width=1.0\linewidth]{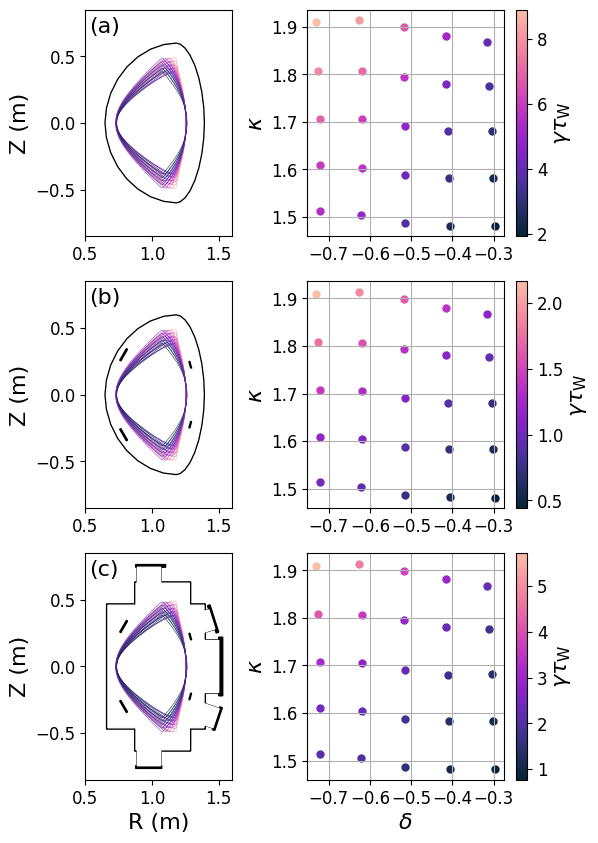} 
    \caption{Feedback capability parameter $\gamma \tau_\mathrm{W}$ across geometry scan for (a) a vacuum vessel conformal to the baseline geometry, approximately 10 cm from the plasma boundary, (b) the same vacuum vessel with copper stabilizing plates, and (c), the final vacuum vessel design with copper stabilizing plates.} 
    \label{fig:vertical_stability}%
\end{figure}

Given that the conducting vacuum vessel is insufficient to mitigate vertical stability, passive stabilizing plates are considered for inclusion in the Negative Triangularity Tokamak. Passive stabilizing plates have been implemented for instability control in existing tokamak experiments such as EAST \cite{east_effect_2012}, KSTAR \cite{kstardesign_2002}, and ASDEX-Upgrade \cite{aug_psl} but have not yet been applied to negative triangularity. Unlike for positive triangularity configurations, modeling indicates that passive stabilizers on the high-field side of the device can reduce vertical instability growth rates for negative triangularity plasmas, resulting in added flexibility when incorporating such stabilizers into real devices \cite{guizzo_assessment_2024}. A study of copper plate locations along the plasma boundary for the $\delta$/$\kappa$ scan found that the plates depicted in figure~\ref{fig:vs-b} result in the greatest improvements in vertical stability across the scan while remaining compatible with ports and other machine constraints. Including both the low field side and high field side plates, $\gamma \tau_\mathrm{W}$ decreases by approximately 75\% for most plasma geometries.

When including the optimized passive plates with the original vacuum vessel, all vertical instability growth rates are finite, and most geometries are in a reasonable domain for active control, as depicted in figure~\ref{fig:vs-c}. As such, the passive plates result in significant improvements for vertical stability prospects in the Negative Triangularity Tokamak. The feedback capability parameter still has significant variation over the geometry scan, a desirable feature for testing control software. However, it should be noted that the vacuum vessel is expected to undergo further design iterations in order to optimize for improved vertical stability.

%\begin{figure}
 %   \centering
    %\includegraphics[width=1\linewidth]{vs_noplate_802.png} 
  %  \caption{Feedback capability parameter $\gamma \tau_\mathrm{W}$ across geometry scan assuming a vacuum vessel conformal to the baseline geometry, approximately 10 cm from the plasma boundary.} 
  %  \label{fig:vs_noplate}%
%\end{figure}

%\begin{figure}
   % \labelphantom{fig:plate-a}
  %  \labelphantom{fig:plate-b}
  %  \centering
    %\includegraphics[width=1\linewidth]{plates_804.png} 
  %  \caption{(a) Final passive plate design, informed by (b) vertical instability growth rate reduction for the baseline scenario due to an up/down symmetric pair of passive plates at locations along the plasma boundary.} 
  %  \label{fig:plate_scan}%
%\end{figure}

%\begin{figure}
%    \centering
    %\includegraphics[width=1\linewidth]{vs_plate_802.png} 
 %   \caption{Feedback capability parameter $\gamma \tau_\mathrm{W}$ across geometry scan with the passive stabilizing plates depicted in figure~\ref{fig:plate-a}.} 
%    \label{fig:vs_plate}%
%\end{figure}

%%%%%%%%%%%%%%%%%%%%%%%%%%%%%%%%%%%%%%%%%
%%%%%%%%%%%%% SECTION FOUR%%%%%%%%%%%%%
%%%%%%%%%%%%%%%%%%%%%%%%%%%%%%%%%%%%%%%%%
\section{Forces on coils and passive structures}
\label{sec:force}
%Forces during standard operation inform required structural supports
% Can approximate current distribution in TF, compute toroidal pressure on TF -> overturning force
% JxB force on PFs computed the same way, integrating over domain on mesh
%Large peak forces, but minimal net radial/vertical forces on PFs
%important to understand disruption loading on passive structures
%perform nonlinear time dependent equilibrium evolution
%to simulate current quench, linearly scale down plasma current over a period of 0.65 ms
%Compute eddy currents induced in passive structures during and after the quench
%shows snapshots, forces concentrated primarily in sharp corners of device -> trade off between ease of construction/placement of PFs and peak forces
%peak net forces are __
% efficacy of passive structures comes from eddy current response -> as a result, particularly susceptible to large disruption forces
%compute radial and vertical force densities on plates
%peak net forces are ___
%these calculations inform further design iterations + design of structural support system for passive structures
% \begin{figure*}
%     \labelphantom{fig:coilforces-a}
%    \labelphantom{fig:coilforces-b}
%    \labelphantom{fig:coilforces-c}
%     \centering
%     \includegraphics[width=1\linewidth]{coilforces_1219.png} 
%     \caption{(a) Pressure on the toroidal field coils, (b) radial force density on the poloidal field coils, and (c) vertical force density on the poloidal field coils for the scenario with maximum currents in all poloidal field coils.}
%     \label{fig:coilforces}%
% \end{figure*}

Both active and passive structural components can be subject to large magnetic forces during both standard tokamak operation and off-normal events \cite{pustovitov_models_2022}. Quantifying such forces is essential for evaluating the feasibility of tokamak designs and informing the required structural supports necessary to prevent device failure and ensure mechanical resiliency.  
%\vspace{12 pt}

The magnetic forces experienced by the toroidal and poloidal field coils can be estimated within the TokaMaker framework. A current of 939~kA per toroidal field coil is required to achieve the desired on-axis magnetic field of 3~T, as discussed further in section~\ref{sec:mag_syst}. The current density is assumed to be uniform across a 2D computational mesh of the toroidal field coil domain.  To predict conservative limiting forces expected for the toroidal field coils, the current in the central solenoid and poloidal field coils are set to their maximum expected value. TokaMaker is then used to evaluate the local magnetic field on the toroidal field coil mesh.

\begin{figure}
    \centering
    \includegraphics[width=1\linewidth]{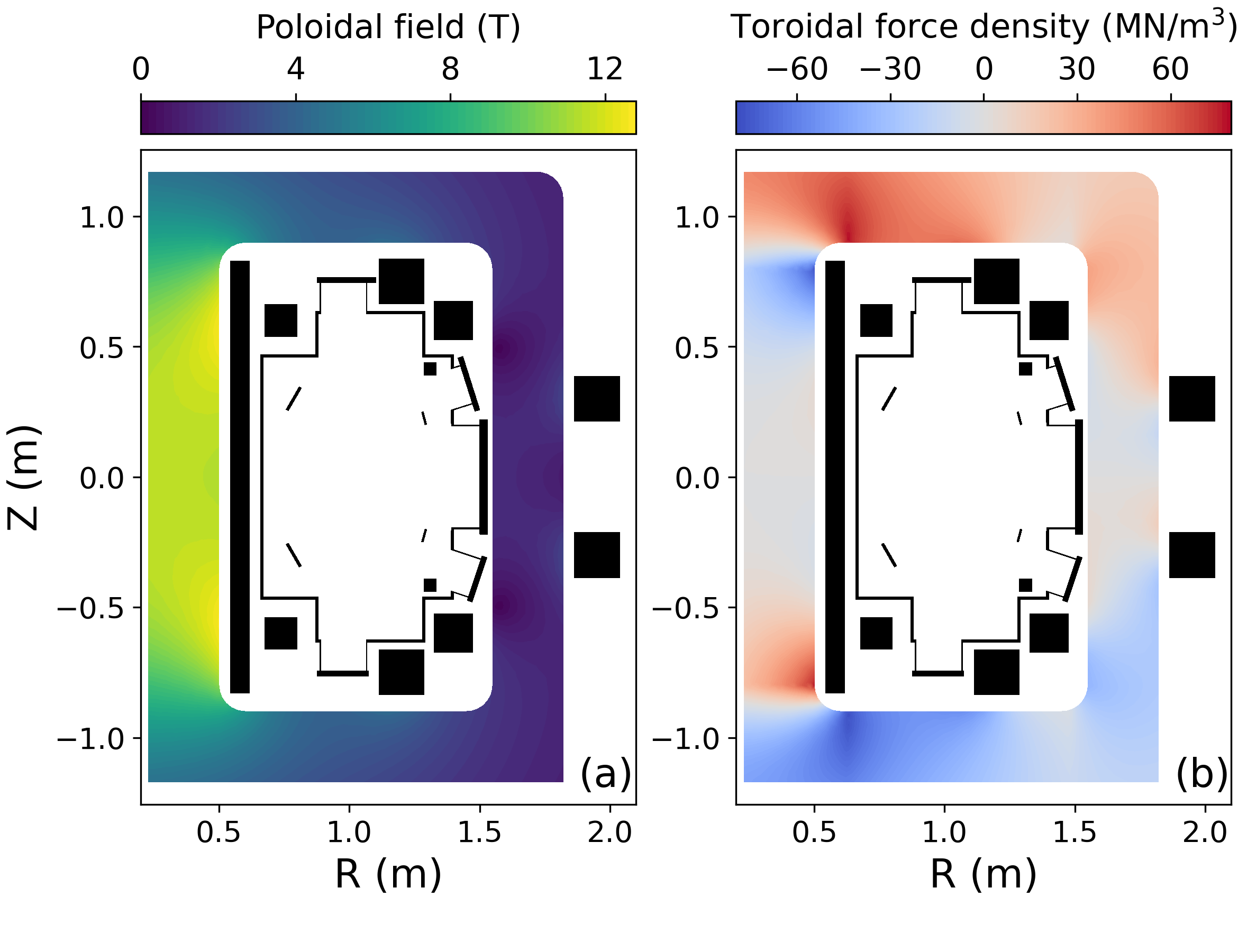} 
    \caption{ (a) Maximum poloidal field and (b) resulting toroidal/overturning force density predicted for NTT using maximum TF/CS/PF coil currents.}
    \label{fig:tf_forces}%
\end{figure}

\begin{figure}
    \centering
    \includegraphics[width=1\linewidth]{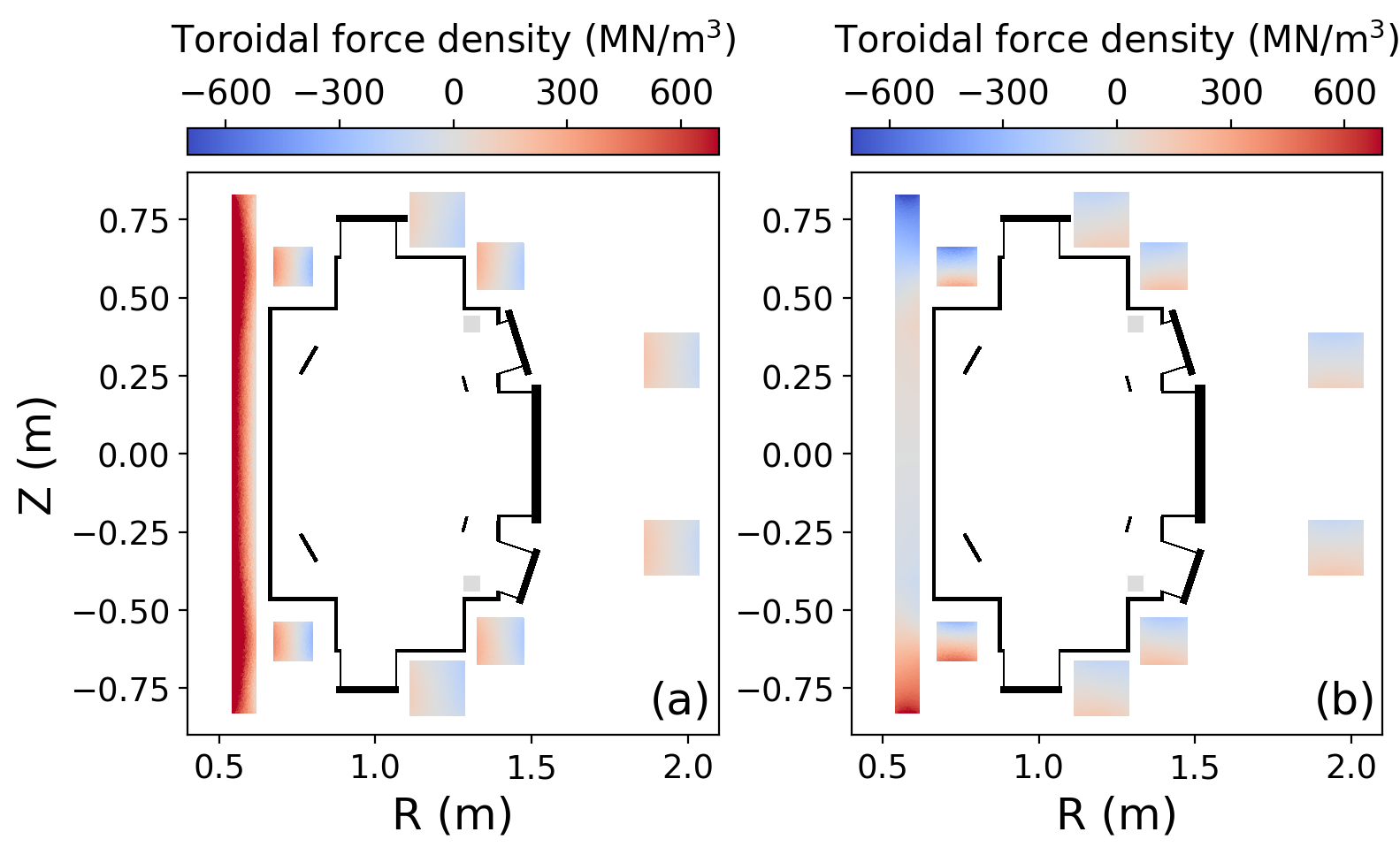} 
    \caption{ (a) Radial and (b) vertical force densities on the poloidal field coils predicted for NTT using maximum CS/PF coil currents.}
    \label{fig:pf_forces}%
\end{figure}

The resultant toroidal force distribution on the coil is shown in figure~\ref{fig:tf_forces}. The up-down asymmetric nature of the toroidal force results in an \enquote{overturning force} on the toroidal field coil which must be overcome with structural supports.  The net torque on a single TF is computed to be $\approx$ 6,000 kN-m.

\begin{figure*}
   \labelphantom{fig:quench-a}
   \labelphantom{fig:quench-b}
   \labelphantom{fig:quench-c}
   \labelphantom{fig:quench-d}
   \labelphantom{fig:quench-e}
   \labelphantom{fig:quench-f}
    \centering
    \includegraphics[width=0.95\linewidth]{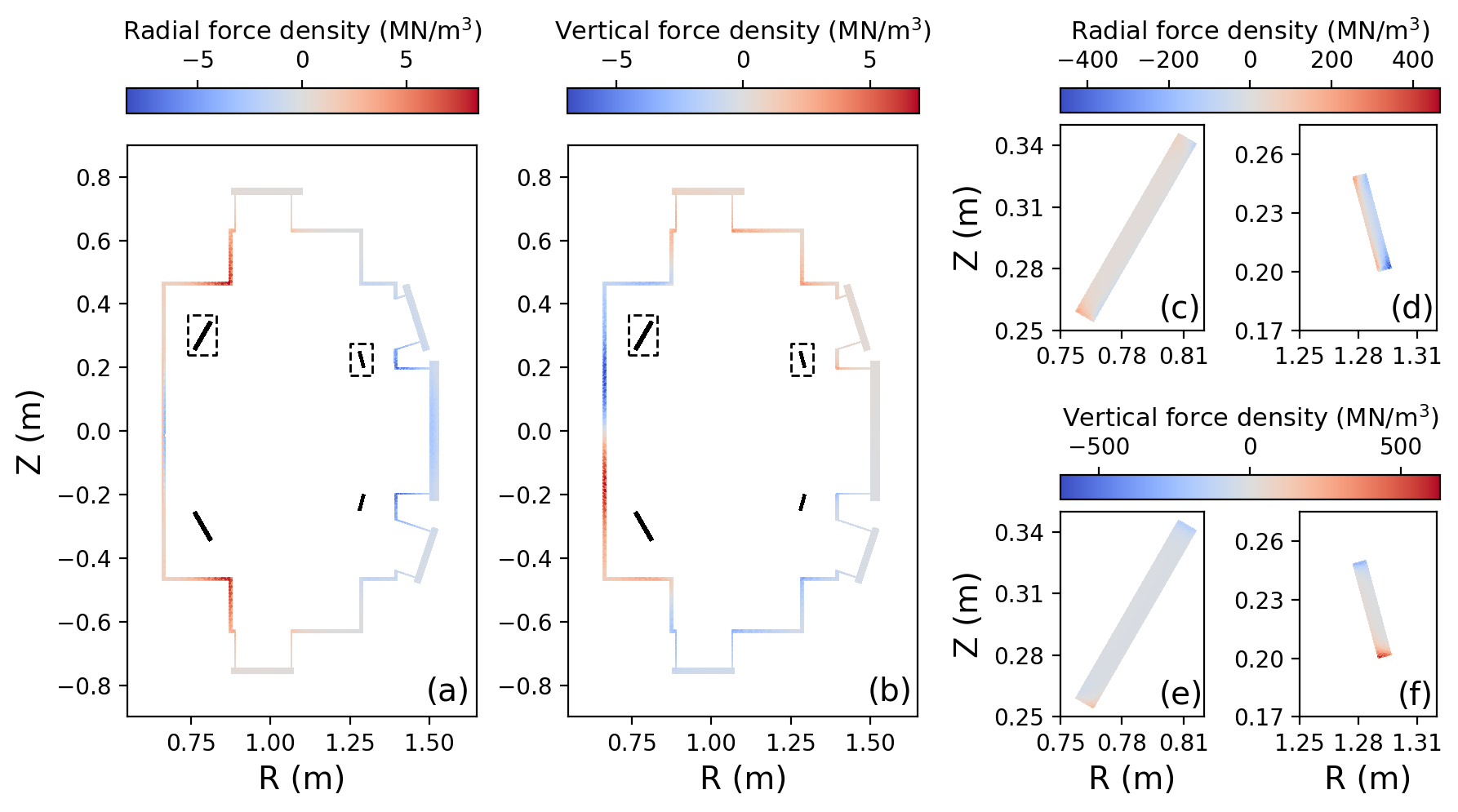} 
    \caption{(a) Radial and (b) vertical force densities on the vacuum vessel, radial force densities on the (c) inner and (d) outer passive plates, and vertical force densities on the (e) inner and (f) outer passive plates, at the end of a 0.65 ms current quench. } 
    \label{fig:quench_forces}%
\end{figure*}

A similar procedure is employed to evaluate the peak forces on the poloidal field coils, for the scenario where each poloidal field coil carries it's maximum current. However, in this case the current in PF3 is reversed, which leads to higher inter-coil forces. The radial and vertical force densities on the PFs for this scenario are plotted in figure~\ref{fig:pf_forces}. Overall, the net radial and vertical forces on each PF are low, with the exception of the CS and PF1, which experience large radially outward force.

The NTT must also be resilient to off-normal events in order to ensure device longevity. In addition to evaluating the forces on active conducting components, we quantify the forces on passive conductors, such as the vacuum vessel and passive stabilizing plates, during a plasma disruption. To simulate a disruption in TokaMaker, the magnetic flux produced by the plasma current is linearly ramped down over a current quench time of 0.65 ms, estimated from the multi-machine scaling given by the ITPA disruption database \cite{eidietis_itpa_2015}. The eddy currents induced in passive conducting structures are evaluated in TokaMaker at each time step, both during and after the quench.  Coupled with the evaluation of the local magnetic field, the resulting magnetic forces on conducting structures are computed as a function of time.

The passive conducting structures experience peak forces at the end of the current quench. Figures ~\ref{fig:quench-a} and ~\ref{fig:quench-b} show the radial and vertical force densities on the vacuum vessel at $t = 0.65$~ms in the simulation. High force densities are concentrated in the thin, sharp corners of the vacuum vessel cross-section, which are largely included for ease of manufacturing and compatibility with poloidal field coil locations.  This result highlights the trade-off between vaccum vessel manufacturing simplicity and resilience to disruptions.

Passive stabilizing plates, in particular, can be subject to large disruption forces.  The considerable eddy-current response in highly conductive passive plates, making them effective in stabilizing magneto-hydrodynamic modes, increases the likelihood of structural damage during uncontrolled plasma evolution.  Figures ~\ref{fig:quench-c}, ~\ref{fig:quench-d}, ~\ref{fig:quench-e}, and~\ref{fig:quench-f}  show the radial and vertical force densities on the inboard and outboard passive stabilizing plates presented in section~\ref{sec:vert_stability}. The net forces per meter of toroidal extent are roughly 28 kN/m and 16 kN/m for the inboard and outboard plates, respectively, and the peak force densities on the passive plates are two orders of magnitude higher than those on the vacuum vessel.  As a result, significantly increased structural support will likely be required to secure the passive plates within the machine. 

The simple force calculations executed in TokaMaker and presented in this section inform further design iterations and set the requirements for structural support systems in the Negative Triangularity Tokamak. Future work will involve the application of higher-fidelity solid mechanics tools to device design.

%%%%%%%%%%%%%%%%%%%%%%%%%%%%%%%%%%%%%%%%%
%%%%%%%%%%%%% SECTION FIVE %%%%%%%%%%%%%
%%%%%%%%%%%%%%%%%%%%%%%%%%%%%%%%%%%%%%%%%
\section{Magnetic system preliminary design}
\label{sec:mag_syst}
\subsection{Toroidal field coil system}
\label{sec:tf_coils}

The toroidal field magnetic system, depicted in  figure~\ref{fig:tf_design-a}, consists of 16 coils resulting in a 1\% ripple in the toroidal field at the plasma edge. The target toroidal field value of 3 T is achieved with 313 kA in each coil. The stray field generated by the current transition between the coils is compensated for by the reverse turn. 

\begin{figure}
    \centering
    \labelphantom{fig:tf_structure-a}
    \labelphantom{fig:tf_structure-b}
    \includegraphics[width=0.675\linewidth]{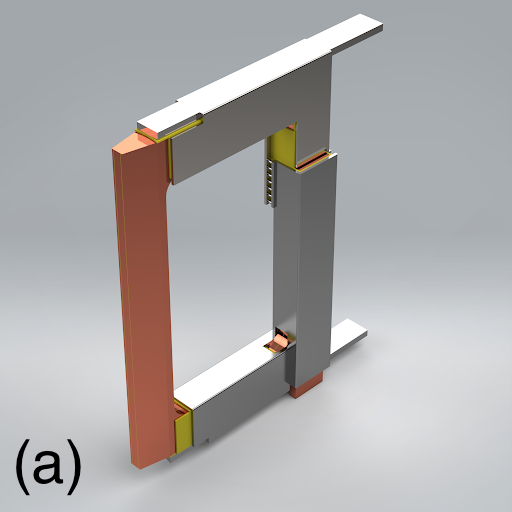} 
    \includegraphics[width=0.675\linewidth]{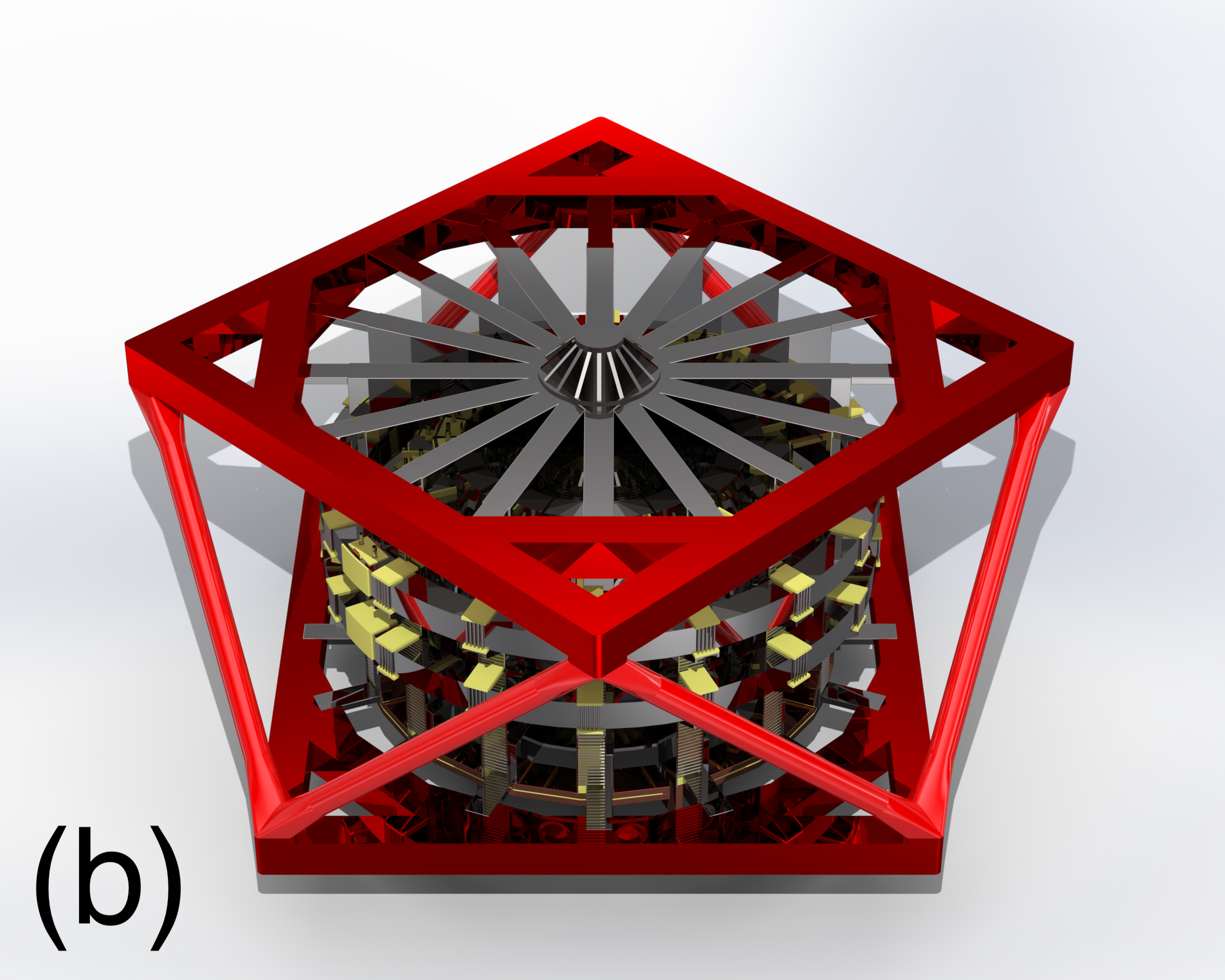} 
    \caption{ (a) Toroidal field coil in steel shell and (b) toroidal field coil shells secured by external support.}
    \label{fig:tf_structure}%
\end{figure}

A single TF coil, shown in figure~\ref{fig:tf_design-b}, is made of three copper turns. The turns are isolated by a thin layer of the polyimide film, resulting in several advantages.
\begin{figure}
    \centering
    \labelphantom{fig:tf_design-a}
    \labelphantom{fig:tf_design-b}
    \includegraphics[width=0.6275\linewidth]{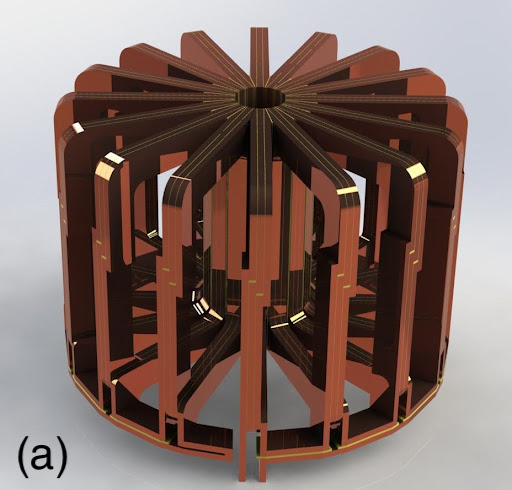} 
    \includegraphics[width=0.3325\linewidth]{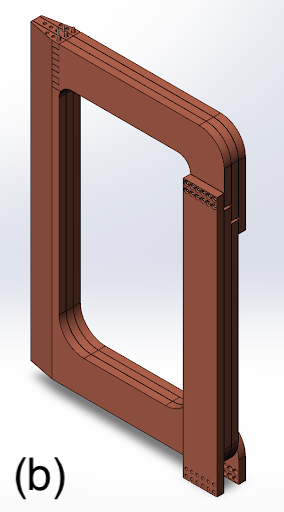} 
    \caption{(a) Full, 16 toroidal field coil assembly without support structure and (b) single three-turn copper toroidal field coil.}
    \label{fig:tf_design}%
\end{figure} 
First, it allows for a high fill factor ($>$0.9), crucial for achieving the target toroidal field of 3 T in the dimensions of NTT. The second advantage is the low resistance and inductance of the coil, reducing the power supply voltage and its complexity. The third advantage of the TF coil design is the higher load-bearing capacity of the turns. The TF coil still needs a metal case for additional stiffness and an outer frame to overcome the overturning force, as shown in figure~\ref{fig:tf_structure}. The total deformation at maximum load does not exceed 5 mm.

\subsection{Poloidal field system}

As mentioned in Section ~\ref{sec:pf_coils}, the TF coils are demountable and have two joints (one on the high-field side and one on the low-field side), allowing for the extraction of the PF coils and the vacuum vessel. 75\% of a turn is made from a single copper plate. During discharge, the coil heats up to 100\textcelsius. The cooling channels are milled into the body of each turn. The required flow of water is $5\cdot10^{-2}~\text{m}^3/\text{s}$ for all 16 TF coils, which can be provided using a standard pressure of 5 atmospheres. Using this system, the TF coils can be cooled down in less than 15 minutes. The total mass of the copper winding of the toroidal magnetic system is about 45 tons.

As mentioned in section~\ref{sec:pf_coils}, the NTT poloidal magnetic system consists of four vertically symmetric pairs of poloidal field coils (PF1-4) and a central solenoid (CS) divided into three sections, as depicted in figure~\ref{fig:cross_section}. All PF coils and CS sections are manufactured as single-layer flat copper busbar coils without integrated cooling channels. 

\begin{figure}
    \centering
    \labelphantom{fig:PF_design-a}
    \labelphantom{fig:PF_design-b}
    \includegraphics[width=0.75\linewidth]{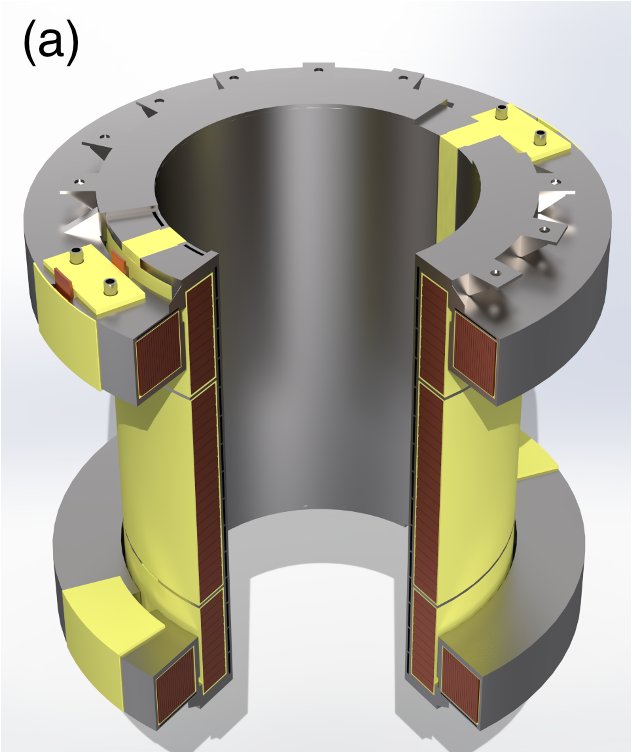} 
    \includegraphics[width=0.75\linewidth]{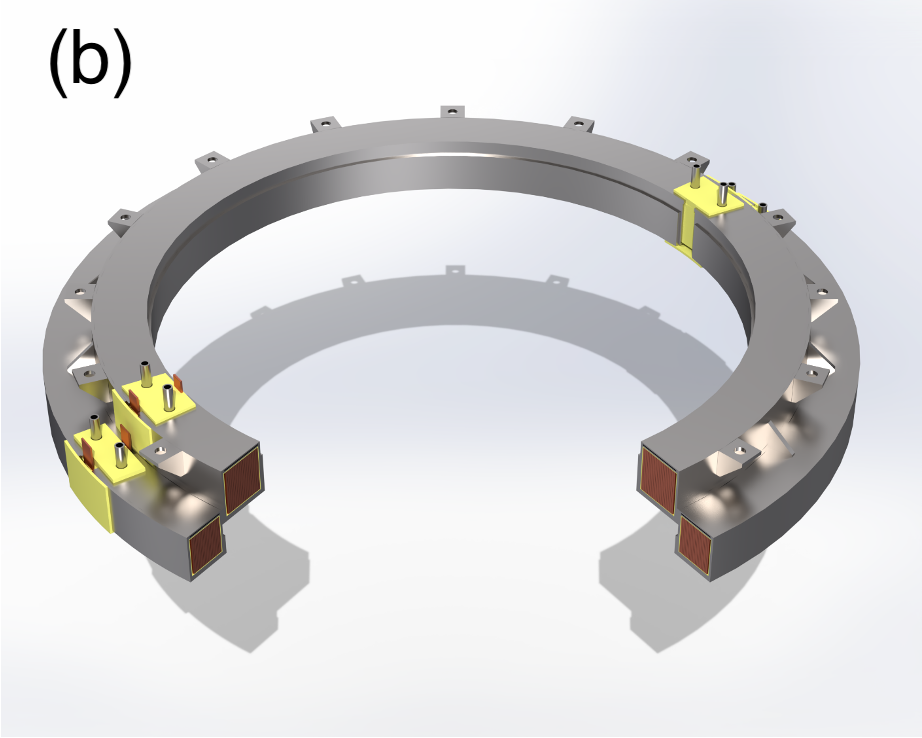} 
    \caption{(a) Preliminary design of the central solenoid assembly combined with PF1 coils and (b) preliminary design of PF2-3 coils assembly.}
    \label{fig:PF_design}%
\end{figure}

This design reduces the resistance and inductance of the coils and increases the fill factor to 0.9. Cooling is achieved by flowing water through wide channels located on the top side of the PF coils and the inner side of the CS. The cooling channel has low hydrodynamic resistance and allows for high water flows. The coil cases are divided into two halves to prevent toroidally closed eddy currents. Each half has its own inlet and outlet for cooling. The coil feeders are located between the two halves of the case. The temperature increase of the PF coils during a discharge is estimated to be less than 70 degrees Celsius, which can be cooled in about 15 minutes. The total weight of the copper coils of the poloidal magnet system is 22 tons. 

Due to the proximity of the PF1 coil and CS, they were combined into one assembly. The preliminary design of the CS and PF1 assembly is shown in figure~\ref{fig:PF_design-a}. The central solenoid assembly will be mounted inside the TF coils and rest on their cases. The PF2 and PF3 coils are also located close to one another and have strong currents in opposite directions, resulting in high mechanical loads between these coils. Therefore, they are combined into a single assembly with additional stiffening ribs, as depicted in figure~\ref{fig:PF_design-b}. PF2-3L assembly will rest in the TF coil cases similar to the CS assembly. The PF2-3U assembly will be attached to the TF coil cases using special clamps. The PF4 coils are designed similarly and are attached to the outer side of the TF coil using clamps.

%%%%%%%%%%%%%%%%%%%%%%%%%%%%%%%%%%%%%%%%%
%%%%%%%%%%%%% SECTION SIX %%%%%%%%%%%%%
%%%%%%%%%%%%%%%%%%%%%%%%%%%%%%%%%%%%%%%%%
\section{Performance analysis}
% \addcontentsline{toc}{section}{VI. Performance Analysis}
\label{sec:perf_analysis}

 While the primary focus of this work concerns the conceptual design of the electromagnetic system for the NT device, it is important to note that many of the fundamental design decisions described above are driven by plasma physics and performance targets. Building off of similar work for the SPARC, ARC, and MANTA machines \cite{creely_2020, sorbom_arc_2015, collaboration_manta_2024}, we adopt a power-balance approach to assess the probable plasma performance of this design. By combining models of the plasma profiles and machine parameters with zero-dimensional scaling laws, a series of Plasma OPerational CONtours (POPCONs) \cite{houlberg_contour_1982} are generated and used to explore the operational range of the target machine. From this initial scoping, the main machine parameters (major radius, minor radius, and elongation) of the baseline scenario depicted in figure~\ref{fig:delta_scan-b} are chosen in order to match the performance objectives of the device.

\begin{figure}
    \centering
    \includegraphics[width=1\linewidth]{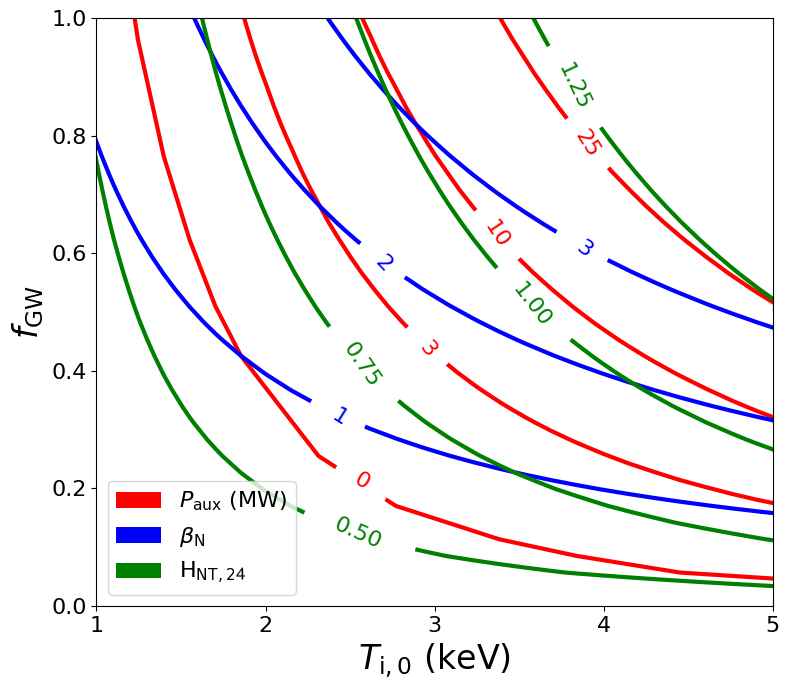} 
    \caption{Plasma OPerational CONtour for the Negative Triangularity Tokamak using the $H_{98, y2} = 1$ confinement scaling.}
    \label{fig:popcon}%
\end{figure}

The results from the POPCON modeling are presented in figure~\ref{fig:popcon}. For this POPCON, 
 the ITER $H_{98,y2}$ energy confinement time scaling relationship is used, setting the confinement quality prefactor $H_{98,y2} =1 $. As depicted in the figure, central ion temperatures up to 2.5 keV can be achieved at $f_{\mathrm{GW}} = 0.2$ with only Ohmic heating. Additional axillary heating of 3 MW allows central ion temperatures as high as 3 keV at $f_{\mathrm{GW}} = 0.45$
 
The calculations performed here are closely informed by previous NT experiments, performed primarily on the DIII-D tokamak \cite{paz-soldan_simultaneous_2024, nelson_robust_2023}. The extensive experimental scans on DIII-D demonstrated that ELM-free operation can be achieved regardless of discharge conditions while simultaneously maintaining normalized betas in the range of $2<\beta_\mathrm{N}\lesssim3$, normalized confinements in the range of $0.5<H_\mathrm{98y2}<1.5$ and safety factors of $2.5\lesssim  q_\mathrm{95}\lesssim4$. As such, the confinement quality factor used to inform the POPCON scoping, as well as the values of $\beta_{\mathrm{N}}$ depicted across the parameter space, should be realizable in a dedicated negative triangularity experiment.

Furthermore, to account for the unique confinement properties of NT plasmas, a specialized NT scaling law was adopted from the DIII-D dataset, as described in \cite{paz-soldan_simultaneous_2024}, for use in predictive power-balance modeling. Compared to the ITER $H_{98,y2}$ scaling, the NT scaling law shows stronger positive correlations with plasma current and average electron density.  However, the NT scaling law also shows stronger energy confinement time degradation with increasing input power. Contours for the $H_\mathrm{NT,24}$ confinement prefactor for the NT scaling law are plotted in figure~\ref{fig:popcon}. For most of the parameter space, the $H_\mathrm{NT,24}$ confinement quality factor needed to reach a given operational point is less than the assumed $H_{98,y2} =1$ confinement quality. This is not true at higher auxillary powers, where the increased power degradation in the NT scaling law results in required confinement quality factors greater than 1.  Inclusion of the NT scaling law analysis increases confidence that the desired operational space can be accessed in a negative triangularity  device.

%As such, the POPCON analysis presented here is \color{red} subject to the following constraints: \color{orange} 

This preliminary performance analysis suggests that the targeted NT tokamak design will be able to reach plasma conditions suitable for the development and testing of sophisticated control systems. Before machine operation, higher-fidelity modeling of the core plasma scenario should be conducted, though this is not expected to significantly impact any of the design decisions described above.
%%%%%%%%%%%%%%%%%%%%%%%%%%%%%%%%%%%%%%
%%%%%%%%%%%%% CONCLUSION %%%%%%%%%%%%%
%%%%%%%%%%%%%%%%%%%%%%%%%%%%%%%%%%%%%%
\section{Summary and Outlook}
\label{sec:conc}
%this work demonstrates feasibility of experimental NT reactor of relevance for fusion startups
%
% leverages tools available in TokaMaker grad shafranov solver to participate in PF design, stability assessments, and disruption loading to give comprehensive overview of the device
%future work: higher fidelity assessment of plasma core scenario, time dependent start-up and ramp down simulations

The Negative Triangularity Tokamak (NTT) proposed in this work is a flexible, compact device dedicated to piloting advanced simulation and control software and informing pilot plant operation.  The NTT is targeting a wide range of plasma geometries with $-0.7 < \delta < -0.3$ and $1.5 < \kappa < 1.9$, providing the ideal test-bed for piloting commercial control software in an adaptive, but reactor-relevant environment.  This study leverages the many tools available in the open-source TokaMaker code \cite{hansen_2023} to design the poloidal field coil system, conduct stability assessments, and predict disruption loads, providing a comprehensive overview of the proposed device.  This work highlights the importance of full-device modeling when designing tokamaks with direct relevance for fusion energy systems.

Future work will build upon the existing design, using the time-dependent capabilities available in TokaMaker to model full plasma scenarios, from start-up to break-down.  Active control of vertical instability will also be modeled in TokaMaker in order to assess the control capabilities of the existing PF coilset and improve upon it where necessary.

In addition to ongoing work with TokaMaker, higher fidelity models will be applied to analyze the NTT core scenario, including the development of heating and fueling scenarios.  Further analysis of the electromagnetic stresses on the device will be conducted by implementing a full 3D model of the conducting structures in the ThinCurr code \cite{hansen_thincurr}, an electromagnetic modeling tool in the OpenFUSIONToolkit.

\section*{Acknowledgements}

This work was supported by Next Step Fusion. S. Guizzo was also supported by Columbia University internal funds.

\label{sec:ack}

  \bibliographystyle{elsarticle-num} 
  \bibliography{mybibfile}



%\begin{thebibliography}{00}

%% For numbered reference style

%\bibitem{lamport94}
%  \textit{\LaTeX: a document preparation system},
%  Addison Wesley, Massachusetts,

%\end{thebibliography}
\end{document}